\documentclass{aa}  

\usepackage{graphicx}
\usepackage{txfonts}
\begin{document} 

   \title{High-velocity CP2 stars in the Galactic halo}

   \author{N. Faltov{\'a}\inst{1} \and M. Pri{\v s}egen\inst{2,1} \and
   K.~Bernhard\inst{3,4} \and S.~H\"ummerich\inst{3,4} \and E. Paunzen\inst{1}}

      \institute{Department of Theoretical Physics and Astrophysics, Faculty of Science, Masaryk University, Kotl\'{a}\v{r}sk\'{a} 2, 611\,37 Brno, Czech Republic
              \and Advanced Technologies Research Institute, Faculty of Materials Science and Technology in Trnava, Slovak University of Technology in Bratislava, Bottova 25, 917 24 Trnava, Slovakia
              \and
        Bundesdeutsche Arbeitsgemeinschaft f\"{u}r Ver\"{a}nderliche Sterne e.V. (BAV), Berlin, Germany
        \and
        American Association of Variable Star Observers (AAVSO), Cambridge, USA \\
              \email{nikola.faltova@gmail.com}
             }

   \date{Received TBA; accepted TBA}

 
  \abstract
   {The second subclass of chemically peculiar stars, the CP2 stars, are early-type stars exhibiting anomalous abundances with variable line strengths possibly also accompanied by photometric variability that typically belong to the Galactic disk. However, a small fraction of these objects were recently found to be located far from the Galactic plane and are thought to belong to the Galactic halo, which is unexpected for this class of objects.}
   {Our study   investigates the halo membership of the proposed halo CP2 star candidates based on their velocities and Galactic orbits, to determine their points of origin. In addition, we searched for further halo CP2 star candidates using an as-yet-unpublished sample of CP stars.}
   {Our analysis relied on the astrometry from the early third \textit{Gaia} data release and radial velocities based on the spectroscopy from LAMOST and SDSS, which was also employed in spectral classification. The light variability of the confirmed CP2 stars in our sample was analyzed using data from the ZTF and ATLAS surveys.}
   {After filtering our initial sample using kinematic and spectroscopic criteria, we identified six CP2 stars with kinematical properties consistent with a halo membership. The orbits of these stars are in agreement with an origin in the Galactic disk where they were probably ejected through dynamical interactions or in the binary supernova scenario, making them the first runaway CP2 stars known.}
   {}


   \maketitle
%

\section{Introduction}

The chemically peculiar (CP) stars are members of the upper main sequence (spectral types ranging from early B to early F), and are notable for the presence of certain absorption lines in their spectra with abnormal strengths, either enhanced or weakened, that indicate peculiar surface abundances. Numerous classes of CP stars have been described; the main groups are the CP1 stars (the metallic-line or Am/Fm stars), the CP2 stars (the Bp/Ap stars), the CP3 stars (the mercury-manganese or HgMn stars), and the He-peculiar stars that encompass the CP4 stars (the He-weak stars) and the He-rich stars \citep[e.g.][]{preston_1974}.

Relevant to this study is the subclass of CP2 stars,   which are known for abnormal abundances and  nonuniform distributions of Si, Sr, Cr, Eu, and other rare earths, and show evidence of strong global magnetic fields of up to several tens of kilogauss  \citep[e.g.,][]{preston_1974,saffe2005,mathys_2017}. Their atmospheres usually exhibit surface abundance patches or spots \citep{michaud_1981,krticka_2013}, leading to photometric 
variability, which is thought to be caused by rotational modulation, and is explained in terms of the oblique rotator model \citep{stibbs_1950}. Within this
model it is essential that the star and the magnetic field are not axially symmetric with respect to the rotation axis. As a result, the observed photometric period is the rotational period of the star. The light curve morphology (shape and amplitude of the light variations) depends on the characteristics of the spots and the line-of-sight conditions \citep{jagelka_2019}. Photometrically variable CP2 stars are known as $\alpha^2$ Canum Venaticorum or ACV variables \citep{GCVS}. For convenience, CP2 stars are often lumped together under the term magnetic chemically peculiar stars (mCP stars) with the He-peculiar stars, which are also thought to harbor strong and organized magnetic fields.

Several studies of different available photometric time series archives that contained a statistically sound sample of CP2 stars were published in recent years
\citep{bernhard_2015a,bernhard_2015b,huemmerich_2016,huemmerich_2018,bernhard_2020,2021MNRAS.506.4561B}. They lead to the
conclusion that the angular momentum during the  main-sequence evolution is conserved; no signs of additional magnetic braking were detected. The inclination angles of the 
rotational axes are randomly distributed, although an apparent excess of fast rotators
with comparable inclination angles was observed \citep{netopil_2017}.

Most CP2 stars tend to be of spectral type B or A, with lifetimes typically of a few hundred Myr. In such a limited time these stars are not expected to move far away from their birthplaces, unless they have been imparted with runaway velocities by some event. Therefore, it came as a surprise that \citet{huemmerich_2020} identified two likely halo CP2 stars in their search for mCP stars in the archives of the Large Sky Area Multi-object Fiber Spectroscopic Telescope \citep[LAMOST;][]{zhao_2012,cui_2012}. Using the astrometric information from \textit{Gaia} data release~2 \citep[GDR2;][]{gaia_mission,gaia_dr2_summary,gaia_dr2_astrometry} and radial velocities from LAMOST \citep{cui_2012}, they concluded that LAMOST J122746.05+113635.3 and LAMOST J150331.87+093125.4 have distances from the Galactic plane and kinematic properties consistent with halo stars. 

In this study we conduct a more detailed investigation of these two stars and the other objects with a height greater than 1200 pc above the Galactic plane from \citet[][see their Table 4]{huemmerich_2020} using updated astrometry from \textit{Gaia} early data release 3 \citep[EDR3;][]{gaia_edr3_summary,gaia_edr3_astrometry} in order to confirm their halo membership and their origins. This sample is supplemented by objects from the unpublished sample of CP stars by Paunzen et al. (in prep.) identified using spectra from the Sloan Digital Sky Survey \citep[SDSS;][]{blanton_2017,Abdurro_2022}.

Baade’s original concept of a stellar population in the Milky
Way \citep{1958RA......5....3B}
was based on a disk (I) and a halo (II) population. The concept later evolved into a scheme with subdivisions. In one such
subdivision, the old metal-weak Population II was divided into
halo (extreme) and intermediate population II objects. The latter was comprised of objects that 
had a velocity dispersion, a chemical composition, and
a concentration toward the Galactic plane that was intermediate between those of the halo and the disk populations. The 
intermediate population II was later proposed in order to explain and fit both the
global and the local Galactic kinematics.
However, \citet{1964ApNr....9..333S} had already pointed out that, owing to a steady transition between the
populations and also to insufficient precision in their ages, no
clear distinction between the groups could be made.

Population II main-sequence A-type stars are born in the Galactic halo, and are therefore intrinsically metal poor. They should not be confused with the blue horizontal-branch stars that occupy the same effective temperature range and were first unambiguously detected and defined in Galactic globular clusters \citep{1962ApJ...135..311A}. 
These stars burn helium in their core and possess a hydrogen-burning shell and a radiative envelope.

The detection of CP stars originating in the Galactic halo would be of great importance for the theories explaining this phenomenon, in particular the origin of the magnetic fields and the efficiency of atomic diffusion within a calm stellar atmosphere \citep{2019MNRAS.482.4519A}. These effects have only been well studied
in a Population I environment. The detection of CP stars of other stellar populations would significantly
improve the astrophysical framework in which the above-mentioned mechanisms are working.

The paper is structured as follows. In Section 2 we describe the data used. In Section 3 we specify the process of the pre-selection of halo candidates and compute the kinematic properties of the pre-selected stars. We filter the sample further and discuss the spectral and photometric properties of the proposed halo objects in Sections 4 and 5, respectively. In Section 6 we compute the Galactic orbits for the retained objects and discuss their properties, and finally provide a brief summary and conclusions in Section 7.

\section{Data}
Studying the kinematical properties of the CP star sample requires phase space information in 6D. \textit{Gaia} is a space-based mission focused on obtaining precise astrometry (positions, parallaxes, and proper motions) for more than a billion sources. The mission also collects other data, most importantly broadband photometry in the $G$, $G_{\mathrm{BP}}$, and $G_{\mathrm{RP}}$ bands. The most recent astrometry data release is the early installment of the third \textit{Gaia} data release \citep[Gaia EDR3;][]{gaia_edr3_summary,gaia_edr3_astrometry} that contains positions, parallaxes, and proper motions for almost 1.5 billion sources, with notable improvements over the previous \textit{Gaia} data release 2 \citep[GDR2;][]{gaia_dr2_summary,gaia_dr2_astrometry} in both formal uncertainties and systematic effects.\footnote{The full Gaia DR3 release has added new radial velocities, astrophysical parameters for a subset of stars, and catalogs of variable and binary stars. However, as both releases are based on the same number of input observations, the astrometry does not change from EDR3 to full DR3.} Therefore, 5D information (spatial position in 3D and 2D velocity in the plane of the sky) can be provided by the \textit{Gaia} mission. 

While \textit{Gaia} also hosts a spectrometer that measures radial velocity, it only does so for a limited subset of relatively bright late-type stars. The stars from the CP sample therefore do not have radial velocities listed in the \textit{Gaia} catalog. For this reason, it is necessary to supplement the \textit{Gaia} astrometry with radial velocity data from ground-based spectroscopic surveys. One such survey is being conducted by LAMOST, which is a four-meter quasi-meridian reflective Schmidt telescope with a field of view of 5$^{\circ}$ \citep{zhao_2012,cui_2012}. It can obtain 4000 spectra in a single exposure at the resolution R$\sim$1800 in the wavelength range of 3700~\AA $<$~$\lambda$~$<$ 9100~\AA. This design makes it most suitable for spectroscopic surveys. LAMOST data products are periodically released to the public in consecutive data releases which can be accessed via the LAMOST archive.\footnote{\url{http://www.lamost.org/lmusers/}} The most recent public data release is DR8, which contains more than 16 million spectra.

In addition, we utilized data from the SDSS, which is a major spectroscopic survey that has been observing the sky for more than 20 years. Since 2017, SDSS has surveyed both hemispheres, observing from Las Campanas Observatory using the du Pont Telescope and the Sloan Foundation 2.5~m Telescope at Apache Point Observatory. The most recent data release is DR17 \citep{Abdurro_2022},\footnote{\url{https://www.sdss.org/dr17}} which is the fifth and final annual release from SDSS-IV.

\section{Identifying halo candidates}

Stars in the Galactic halo are commonly identified kinematically as the fastest moving stars with respect to the local standard of rest (LSR). To do so, we made use of Gaia EDR3 astrometry supplemented by stellar radial velocities derived from LAMOST and SDSS data.

We started with the ten CP2 stars that were identified by \citet{huemmerich_2020} as potential halo objects based on their large distances from the Galactic plane as inferred from GDR2 data (see  in particular their Table 4). We also considered the sample of CP stars identified in SDSS spectra by Paunzen et al. (in prep). The LAMOST sample comprises brighter stars, with most of them being brighter than $G$\,=\,13~mag. The stars in the SDSS sample are significantly fainter; none of them is brighter than $G$\,=\,13~mag.

First, we performed parallax bias correction for all sample stars using the procedure from \citet{gaia_edr3_parallax_correction} via the accompanying code \texttt{gaiadr3\_zeropoint}.\footnote{The package was downloaded from \url{https://gitlab.com/icc-ub/public/gaiadr3_zeropoint}} The exact value of the bias correction depends on the magnitude, color, sky position, and astrometric solution type of the source. In order to get the corrected parallaxes, the bias correction needs to be subtracted from the raw parallaxes listed in the \textit{Gaia} catalog. For all studied stars the bias was negative; therefore, the raw parallaxes were increased. The median increase in the source parallax was 0.039~mas for the LAMOST stars and 0.034~mas for the SDSS stars.

Since six of the sources in the LAMOST sample are brighter than $G$\,=\,13~mag, we also computed the proper motion correction based on the method presented in \citet{cantat-gaudin_2021}. The correction is magnitude- and position-dependent, and is usually quite small and therefore comparable to typical proper motion uncertainties in this magnitude range.

No additional correction was applied to the radial velocities from LAMOST and SDSS. While some effects such as the gravitational redshift and stellar convection can slightly affect the observed radial velocities, these changes are below the sub-$\mathrm{km \, s^{-1}}$ level \citep{gullberg_2002}, which is significantly smaller than the radial velocity uncertainties.

For the initial pre-selection of potential halo objects, we calculated the approximate positions and velocities of the sample stars in the galactocentric frame. The usefulness of astrometry for this science case is usually limited by the accuracy of the parallax. For the kinematics to be sufficiently constrained, we only considered stars with a relative parallax error below $\sim$20--25$\%$ (\texttt{parallax\_over\_error} $>4$). All LAMOST stars fulfill this criterion, but some stars from the SDSS sample had to be removed; 171 CP star candidates were retained.

Complementing the positions, parallaxes, and proper motions from \textit{Gaia} EDR3 with the radial velocities from LAMOST and SDSS provided us with the full 6D phase space information needed to calculate the galactocentric positions and velocities of the studied stars, and to constrain their trajectories back in time to potentially uncover their origin and constrain their ejection velocities. Uncertainty propagation in the parallaxes, proper motions, and radial velocities was done using the Monte Carlo approach with $10^{5}$ realizations, under the assumption that the listed uncertainties are 1$\sigma$ errors of the normal distribution centered on the measured value of the corresponding parameter. Correlations between the parameters provided by \textit{Gaia} EDR3 were taken into   account. We adopted 
\begin{equation}
\label{eq:mean}
\mathbf{m} = [\mu_{\alpha*}, \mu_\delta, \varpi]
\end{equation}
and the covariance matrix
\begin{equation}
\label{eq:covmatr} 
\footnotesize
\arraycolsep=2pt
\thickmuskip =1.mu
\Sigma = {}
\left(
\begin{array}{@{}ccc@{}}
\sigma_{\mu_{\alpha *}}^2 & \sigma_{\mu_{\alpha *}} \sigma_{\mu_\delta} \rho(\mu_{\alpha *},\mu_\delta) & \sigma_{\mu_{\alpha *}} \sigma_\varpi \rho(\mu_{\alpha *},\varpi) \\
\sigma_{\mu_{\alpha *}} \sigma_{\mu_\delta} \rho(\mu_{\alpha *},\mu_\delta) & \sigma_{\mu_\delta}^2 & \sigma_{\mu_{\delta}} \sigma_\varpi \rho(\mu_\delta,\varpi) \\
\sigma_{\mu_{\alpha *}} \sigma_\varpi \rho(\mu_{\alpha *},\varpi) & \sigma_{\mu_\delta} \sigma_\varpi \rho(\mu_\delta,\mu_\varpi) & \sigma_\varpi^2 
\end{array}
\right).
\end{equation}Here and throughout this work, to compute the  galactocentric positions and velocities (and their uncertainties), we used the right-handed galactocentric frame defined such that the origin is located in the Galactic Center and the X-axis is aligned with the Sun's direction in accordance with the Sun being located at X=-8.122~kpc \citep{gravity_2018}. The Y-axis points approximately  toward Galactic longitude $l=90^{\circ}$ and the Z-axis toward the north Galactic pole ($b=90^{\circ}$). The 3D solar velocity is fixed to $\mathbf{v_{\odot}} = [12.9, 245.6, 7.78]$~$\mathrm{km \, s^{-1}}$ \citep{drimmel_2018}. The height of the Sun above the Galactic disk is assumed to be 20.8~pc \citep{bennett_2019}.

\begin{figure}
\centering
\includegraphics[width=\hsize]{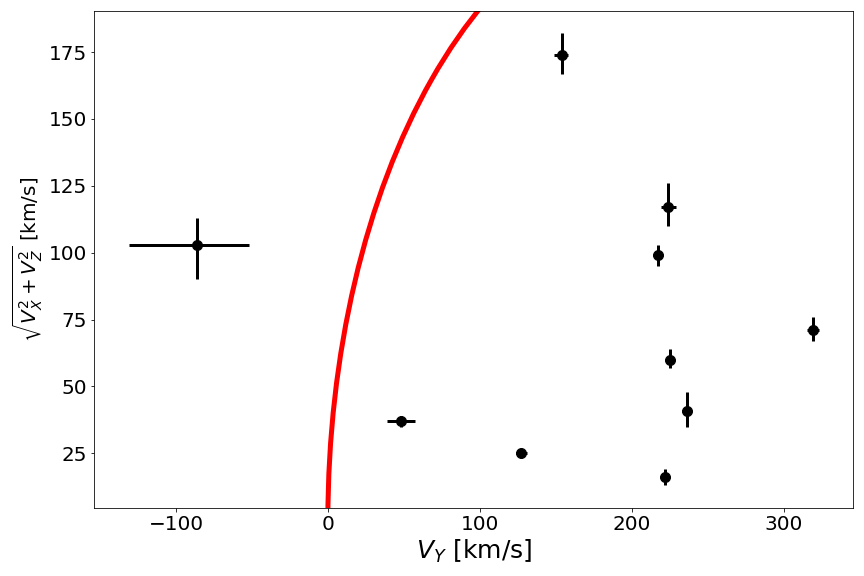}
\includegraphics[width=\hsize]{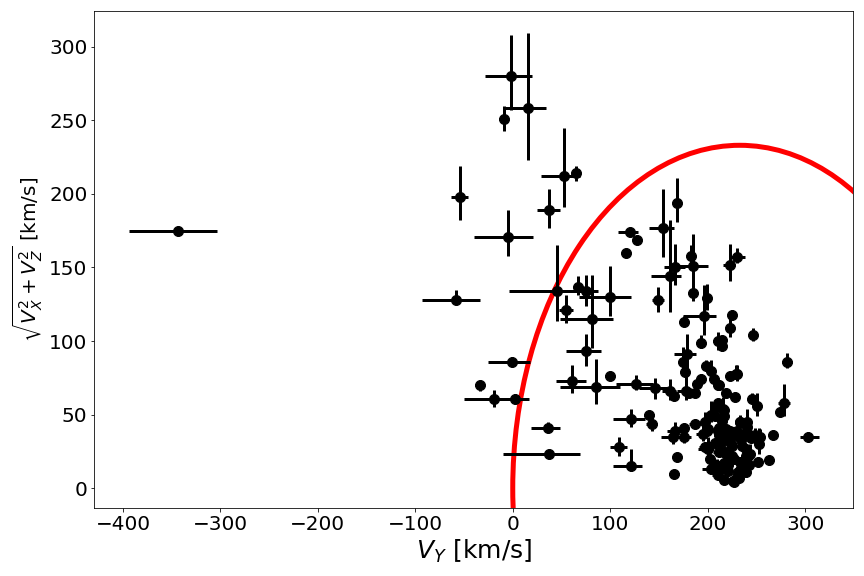}

\caption{Toomre diagrams for the studied stars from the LAMOST sample (top) and SDSS sample (bottom). The red line represents the boundary between the halo and disk stars, as given in \citet{bonaca_2017}.}
\label{toomre_diagram}
\end{figure}
Figure \ref{toomre_diagram} shows a Toomre diagram for the studied stars, which can be used to distinguish stars belonging to different Galactic populations (thin disk, thick disk, and halo) according to their kinematic properties. The galactocentric velocity $V_{Y}$, which is the velocity of an object along the direction of the Galactic rotation, is plotted on the x-axis. The velocity component orthogonal to that (i.e., $\sqrt{V_{X}^{2} + V_{Z}^{2}}$) is on the y-axis. It can be seen that the majority of sources are concentrated near $\sim$(230, 0)~$\mathrm{km \, s^{-1}}$, which is consistent with the kinematics of the objects belonging to the Galactic disk. In addition that, a more diffuse population with a larger spread in velocities is visible as well. The red line broadly divides the disk and halo components, where the halo stars are selected to have $|V-V_{\mathrm{LSR}}|>233$~$\mathrm{km \, s^{-1}}$ \citep[e.g.,][]{bonaca_2017}, where $V_{\mathrm{LSR}} \approx (0, 233, 0)$~$\mathrm{km \, s^{-1}}$ \citep{drimmel_2018}.

\begin{table*}
\caption{Astrometric data for the candidate halo objects. The star Gaia DR3 3907547639444408064 (LAMOST 122746.05+113635.3) is from the sample of candidate halo CP2 stars from \citet{huemmerich_2020}; all other objects are from the unpublished sample of CP star candidates by Paunzen et al. (in prep.).}             
\label{table1_astrometry}      
\centering          
\begin{tabular}{lllllll}
\hline 
\hline
GDR3 source\_id & RA & Dec &$v_{\mathrm{r}}$ & $\varpi$ & $\mu_{\alpha^{\star}}$ & $\mu_{\delta}$ \\
 & (deg) & (deg) &($\mathrm{km \, s^{-1}}$) & (mas) & ($\mathrm{mas \, yr^{-1}}$) & ($\mathrm{mas \, yr^{-1}}$) \\
\hline
3907547639444408064 & 186.942 & 11.61 & 182$\pm$5 & 0.238$\pm$0.039 & -10.097$\pm$0.046 & -13.825$\pm$0.043 \\
1238987465193757440 & 216.17 & 18.662 & -138$\pm$2 & 0.249$\pm$0.029 & -21.398$\pm$0.034 & 0.02$\pm$0.028 \\
1311569698078118656 & 252.073 & 30.475 & -256$\pm$2 & 0.205$\pm$0.045 & -3.88$\pm$0.052 & -9.319$\pm$0.047 \\
1337578198956516736 & 256.099 & 33.025 & -261$\pm$1 & 0.255$\pm$0.02 & -5.137$\pm$0.02 & 5.8$\pm$0.023 \\
1419257997205516416 & 260.284 & 53.712 & -279$\pm$4 & 0.234$\pm$0.028 & -9.308$\pm$0.038 & 8.67$\pm$0.038 \\
1419489096510038272 & 259.463 & 53.901 & -267$\pm$2 & 0.307$\pm$0.022 & -8.183$\pm$0.025 & 1.173$\pm$0.029 \\
1427740179658449536 & 242.397 & 52.153 & -147$\pm$2 & 0.169$\pm$0.031 & -1.872$\pm$0.034 & -12.049$\pm$0.04 \\
1523216134417639040 & 198.328 & 39.036 & -6$\pm$1 & 0.259$\pm$0.03 & -9.796$\pm$0.023 & -13.158$\pm$0.026 \\
1525084891867813888 & 201.344 & 40.902 & -137$\pm$2 & 0.227$\pm$0.036 & -1.271$\pm$0.023 & -14.215$\pm$0.03 \\
1546648032878846336 & 182.25 & 49.65 & -119$\pm$1 & 0.294$\pm$0.024 & -9.838$\pm$0.014 & -11.772$\pm$0.019 \\
3675859515607921664 & 191.618 & -7.268 & 166$\pm$2 & 0.277$\pm$0.053 & -15.501$\pm$0.072 & 1.342$\pm$0.063 \\
3904797657784182784 & 183.892 & 8.159 & 84$\pm$2 & 0.418$\pm$0.031 & -26.123$\pm$0.035 & -6.014$\pm$0.026 \\
4574899937071889408 & 256.396 & 27.351 & -234$\pm$1 & 0.416$\pm$0.019 & -24.762$\pm$0.018 & 8.283$\pm$0.02 \\
629949175496488704 & 151.824 & 22.292 & 292$\pm$2 & 0.273$\pm$0.028 & -15.537$\pm$0.03 & -32.721$\pm$0.029 \\
676074826449368704 & 123.744 & 21.614 & 140$\pm$2 & 0.275$\pm$0.033 & -2.709$\pm$0.036 & -15.031$\pm$0.023 \\
\hline
\end{tabular}
\end{table*}

This selection resulted in retaining one star from the LAMOST sample and 14 stars from the SDSS sample as potential halo object candidates. Their astrometric properties are summarized in Table~\ref{table1_astrometry}. We also checked the quality of the astrometric solution for the retained objects as extreme velocities are often spurious results when bad astrometry is used. According to the recommended astrometric quality flags, all objects seem to be astrometrically ``well behaved'' with no duplicated sources. The renormalized unit weight error (\texttt{ruwe}) parameter, which is also provided in GDR3, is well below 1.4 for all the pre-selected objects, which also indicates that there should not be severe problems with the astrometry. Moreover, the same conclusions are reached by inspecting the values of astrometric \texttt{fidelity} values for these sources provided in \citet{rybizky_2022}. These values are close to, or exactly 1.0, which indicates reliable astrometric solutions.


\section{Spectral classification}
\label{spectral_classification}

Spectral classification was performed in the framework of the refined MK classification system \citep{gray87,gray89a,gray89b,gray94,gray09}. To derive a precise classification and identify peculiarities, the blue-violet ($\lambda\lambda$3800$-$4600\,\AA) spectral regions of the target star spectra were overlaid with and compared to MK standard star spectra from the \textit{libr18\_225} library. This library is distributed with Richard O. Gray's MKCLASS code,\footnote{\url{http://www.appstate.edu/~grayro/mkclass/}} a computer program written to classify stellar spectra on the MK system \citep{gray14}, and consists of standard star spectra smoothed to a reduced resolution of 2.25\,\AA\ that were obtained with the Gray Miller Spectrograph at the 32-inch telescope of Appalachian State University's Dark Sky Observatory.\footnote{\url{https://dso.appstate.edu/}} The spectral classification for Gaia DR3 3907547639444408064 (LAMOST J122746.05+113635.3) was taken from \citet{huemmerich_2020}.

Out of the 15 investigated stars, 12 objects show obvious discrepancies between the spectral types derived from the hydrogen lines (h-line type), the \ion{Ca}{ii} K line (k-line type), and the general strength of the metallic-line spectrum (m-line type). Therefore, where appropriate, spectral types based on these different criteria are provided in the classifications given in Table \ref{table_SpT}. The main peculiarities for the CP2 stars are given in order of importance. Parentheses denote a moderate enhancement; a colon is used to identify uncertain classifications. 

Seven stars show k-line and m-line types earlier than the h-line types, which implies that the stars are metal weak. This is expected for halo stars. There is also one star that is a chemically normal early F dwarf, and another one with apparently the same spectral type but the corresponding spectrum is of very low S/N and the classification has to be regarded as only an estimate. In consequence, these nine stars were excluded from further consideration in this study.

\begin{figure*}
        \includegraphics[width=\textwidth]{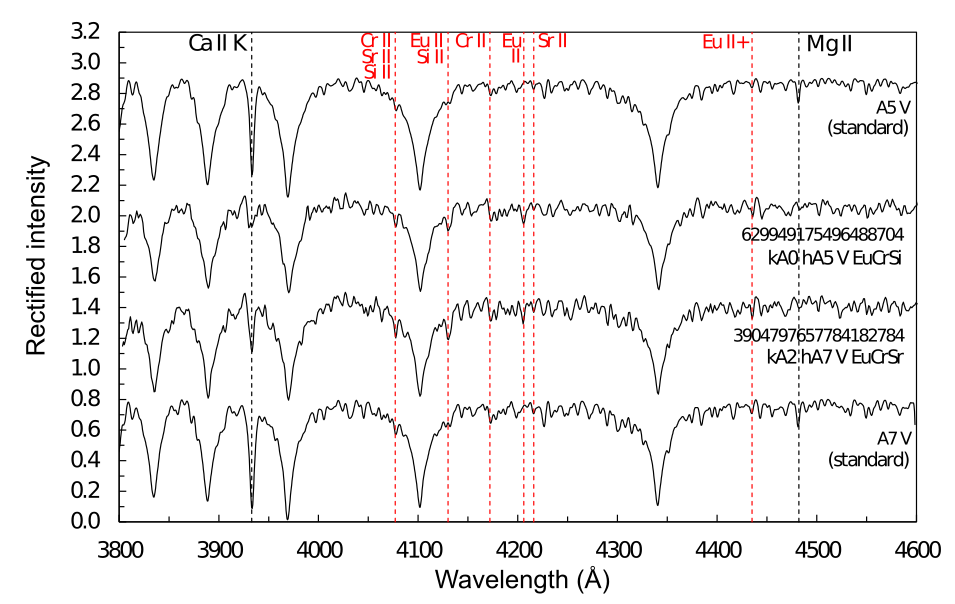}
    \caption{Spectra of two halo CP2 stars in comparison to the A5 V and A7 V standard stars from the \textit{libr18\_225} library. Some prominent lines of interest are identified; lines relevant to the classification of CP2 stars are highlighted in red. There are strong \ion{Eu}{ii} features in the CP2 stars.}
    \label{showcase}
\end{figure*}

The remaining six stars are bona fide CP2 stars. There are five cool CP2 stars, whose peculiar metallic-line spectra are dominated by features of \ion{Eu}{II}, \ion{Cr}{II}, \ion{Sr}{II}, and, to a lesser extent, \ion{Si}{II}, and one hot Si CP2 star (see  Table \ref{table_SpT}). Two example spectra are presented in Fig. \ref{showcase}. Interestingly, except for the hot CP2 star, all these objects show a rather similar abundance pattern that is, at least in four objects, strongly dominated by enhanced \ion{Eu}{II} features. While Eu is commonly enhanced in cool CP2 stars, it rarely is the dominating contributor to the peculiarity mix (see, e.g., \citealt{huemmerich_2020}). It is intriguing to speculate on a possible connection of this characteristic to the unusual kinematic properties of these stars. An investigation of this, however, is beyond the scope of the present study.

\begin{table*}
\caption{Spectral types derived in this study (see Section \ref{spectral_classification} for details). Stars are ordered first by type and then  by increasing right ascension. The classification for Gaia DR3 3907547639444408064 was adopted from \citet{huemmerich_2020}.}
\label{table_SpT}
\centering          
\begin{tabular}{llll}
\hline 
\hline
GDR3 source\_id & ID\_alt & SpT & Type \\
\hline
676074826449368704 & 2MASS J08145859+2136521 & kA3 hF0 V EuSr(Si) & CP2 star \\ 
629949175496488704 & \object{2MASS J10071768+2217330} & kA0 hA5 V EuCrSi & CP2 star \\ 
3904797657784182784 & \object{2MASS J12153411+0809325} & kA2 hA7 V EuCrSr & CP2 star \\ 
3907547639444408064 & \object{LAMOST J122746.05+113635.3} & B8 IV Si (He-wk)$^{(a)}$ & CP2 star \\ 
3675859515607921664 & \object{2MASS J12462821-0716061} & kA0 hA3 V EuSr(Cr) & CP2 star \\ 
1419257997205516416 & \object{2MASS J17210823+5342419} & kA2 hA5: V: SrEuSiCr & CP2 star \\ 
\hline
1546648032878846336 & 2MASS J12085992+4939011 & kA2 hA7 mA5 V & metal-weak \\ 
1523216134417639040 & 2MASS J13131880+3902083 & kA0 hA1 mA0 V & metal-weak \\ 
1525084891867813888 & 2MASS J13252247+4054077 & kA2 hA7 mA2 V (Sr) & metal-weak \\ 
1238987465193757440 & 2MASS J14244070+1839433 & kA2 hA5 mA3 V & metal-weak \\ 
1311569698078118656 & 2MASS J16481749+3028305 & kA2 hA4 mA3 V & metal-weak \\ 
1337578198956516736 & 2MASS J17042374+3301299 & kA3 hA5 mA2 V & metal-weak \\ 
4574899937071889408 & 2MASS J17053507+2721043 & kA2 hA7 mA5 V (Sr) & metal-weak \\ 
\hline
1427740179658449536 & 2MASS J16093537+5209099 & F0 V & non-CP star \\ 
1419489096510038272 & 2MASS J17175123+5354043 & F0: V: & non-CP star? \\ 
\hline
\multicolumn{4}{l}{Notes:} \\
\multicolumn{4}{l}{$^{(a)}$Spectral type from \citet{huemmerich_2020}.} \\
\end{tabular}
\end{table*}


\section{Photometric variability}
\label{photometric_variability}

We examined the photometric variability of the confirmed CP2 stars in our sample using data from the Zwicky Transient Facility (ZTF) and the Asteroid Terrestrial-impact Last Alert System (ATLAS). As the ZTF data are generally superior in terms of precision, where available, the analysis was based on this data source.

The ZTF is a time-domain survey located at Palomar Observatory that employs e2v CCD231-C6 camera devices mounted on the Palomar 48-inch Samuel Oschin Schmidt Telescope. Observations are acquired in three different passbands ($g$, $r$, and $i$) to a limiting magnitude of 20.5 mag. More information on the ZTF survey can be gleaned from \citet{ZTF1} and \citet{ZTF2}. ZTF data have been shown to be ideally suited for the study of the photometric variability of CP2 stars \citep{faltova21}. As only very few data points in $i$ were available, only $g$ and $r$ data were included in our analysis.

ATLAS is a sky survey aiming at the discovery of potentially hazardous near-Earth asteroids (NEAs). It currently employs f/2.0 0.5~m DFM custom Wright Schmidt telescopes and STA-1600 10.5x10.5k CCD detectors that are situated on Haleakala and Mauna Loa. Observations are acquired through two broadband filters: the cyan filter ($c$; 420$-$650\,nm) and the orange filter ($o$; 560$-$820\,nm), which are linked via color transformations to the Pan-STARRS $g$, $r$, and $i$ bands. More information on the ATLAS sky survey is found in \citet{tonry18}.

According to our expectations, we could confirm five of the six CP2 stars as ACV variables with typical amplitudes and periods. Two objects (Gaia DR3 629949175496488704 and Gaia DR3 3904797657784182784) show anti-phase variations between the ZTF $g$ and $r$ light curves:  when the star gets brighter in $g$, it gets fainter in $r$, and vice versa. This is a tell-tale sign of CP2 stars \citep{molnar73,molnar75,groebel17,faltova21} that excludes the possibility that other variability mechanisms, such as pulsation, are at work in these objects. In both stars the null wavelength region  (i.e., the spectral region in which the flux remains almost unchanged over the rotational period) is situated somewhere between 4750 and 6200\,\AA, the central wavelengths of the $g$ and $r$ filters.

Two objects (Gaia DR3 3675859515607921664 and Gaia DR3 1419257997205516416) show variability in the ZTF $r$ band, but no detectable signal in the $g$ light curves, which is also typical of CP2 stars \citep{faltova21}. Gaia DR3 3907547639444408064 shows consistent light curves with similar amplitudes in both ATLAS passbands; no ZTF data are available for this object.

The remaining CP2 star (Gaia DR3 676074826449368704) does not show any consistent variability in either ZTF or ATLAS data. This does not prove that it is not an ACV variable; the amplitude of the light variations may be too small or the period too long to be detected in the employed data sources.

The results from the period analysis are given in Table \ref{table_period_analysis}. Fourier spectra and phased light curves of the CP2 stars are shown in the Appendix in Figs. \ref{fig_Fourier_spectra} and \ref{fig_phase_plots}. Period analysis was performed with the Lomb-Scargle GLS algorithm as implemented in the program package PERANSO \citep{peranso}.

\begin{table}
\caption{Variability periods of the five halo CP2 stars, derived from an analysis of ZTF and ATLAS data with the Lomb-Scargle GLS algorithm. Periods are given to the last significant digit. Data sources are indicated in the remarks.}
\label{table_period_analysis}
\centering          
\begin{tabular}{lll}
\hline 
\hline
GDR3 source\_id & $P$(d) & remark \\
\hline
629949175496488704 & 3.021 & antiphase var. (ZTF) \\
3904797657784182784 & 5.518 & antiphase var. (ZTF) \\
3907547639444408064 & 3.4858 & consistent var. in both \\
 & & $c$ and $o$ (ATLAS) \\
3675859515607921664 & 2.9296 & var in $r$ only (ZTF) \\
1419257997205516416 & 15.83 & var in $r$ only (ZTF) \\
\hline
\end{tabular}
\end{table}


\section{Halo membership and Galactic orbits}

\begin{table*}
\caption{Positions and kinematical properties of the halo CP stars.}             
\label{table2_kinematics}      
\centering          
\begin{tabular}{llllllll}
\hline 
\hline
GDR3 source\_id & $X$ & $Y$ & $Z$ & $V_{\mathrm{X}}$ & $V_{\mathrm{Y}}$ & $V_{\mathrm{Z}}$ & $t_{\mathrm{cross}}$ \\
 & (kpc) & (kpc) & (kpc) & ($\mathrm{km \, s^{-1}}$) & ($\mathrm{km \, s^{-1}}$) & ($\mathrm{km \, s^{-1}}$) & (Myr) \\
\hline
676074826449368704 & ${-10.70}_{-{0.30}}^{+{0.24}}$ & ${-1.01}_{-{0.12}}^{+{0.09}}$ & ${1.48}_{-{0.13}}^{+{0.17}}$ & ${-81}_{-{2}}^{+{3}}$ & ${-1}_{-{23}}^{+{18}}$ & ${-29}_{-{12}}^{+{9}}$ & ${-47.9}_{-{4.3}}^{+{3.9}}$ \\
629949175496488704 & ${-9.77}_{-{0.16}}^{+{0.13}}$ & ${-1.00}_{-{0.10}}^{+{0.08}}$ & ${2.54}_{-{0.20}}^{+{0.24}}$ & ${-173}_{-{3}}^{+{3}}$ & ${-344}_{-{48}}^{+{40}}$ & ${22}_{-{22}}^{+{17}}$ & ${-40.0}_{-{15.3}}^{+{8.9}}$ \\
3904797657784182784 & ${-8.02}_{-{0.01}}^{+{0.01}}$ & ${-0.77}_{-{0.06}}^{+{0.05}}$ & ${2.05}_{-{0.13}}^{+{0.15}}$ & ${-186}_{-{15}}^{+{13}}$ & ${38}_{-{13}}^{+{11}}$ & ${29}_{-{4}}^{+{4}}$ & ${-24.8}_{-{0.8}}^{+{0.7}}$ \\
3907547639444408064 & ${-7.90}_{-{0.03}}^{+{0.04}}$ & ${-1.00}_{-{0.16}}^{+{0.12}}$ & ${3.47}_{-{0.42}}^{+{0.55}}$ & ${-4}_{-{5}}^{+{3}}$ & ${-86}_{-{44}}^{+{35}}$ & ${102}_{-{13}}^{+{11}}$ & ${-24.7}_{-{4.6}}^{+{3.6}}$ \\
3675859515607921664 & ${-7.18}_{-{0.14}}^{+{0.19}}$ & ${-1.56}_{-{0.32}}^{+{0.23}}$ & ${2.68}_{-{0.38}}^{+{0.54}}$ & ${-148}_{-{42}}^{+{30}}$ & ${52}_{-{23}}^{+{17}}$ & ${152}_{-{2}}^{+{3}}$ & ${-14.7}_{-{2.0}}^{+{1.5}}$ \\
1419257997205516416 & ${-7.63}_{-{0.05}}^{+{0.06}}$ & ${3.19}_{-{0.31}}^{+{0.38}}$ & ${2.25}_{-{0.22}}^{+{0.26}}$ & ${-197}_{-{21}}^{+{17}}$ & ${-53}_{-{9}}^{+{7}}$ & ${-9}_{-{14}}^{+{16}}$ & ${-34.0}_{-{0.9}}^{+{1.1}}$ \\
\hline
\end{tabular}
\end{table*}

\begin{figure}
\centering
\includegraphics[width=\hsize]{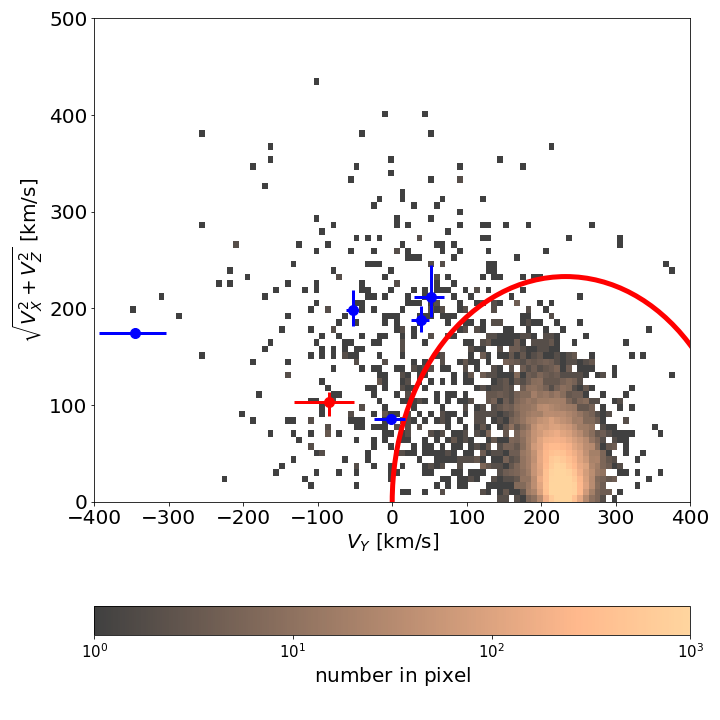}

\caption{Toomre diagram of stars spanning spectral classes from B5 to A5 with reliable astrometry and radial velocity from the catalog of \citet{xiang_2022}. Blue points correspond to the halo CP2 stars identified in the SDSS data and the red data point corresponds to the CP2 star identified in the LAMOST data.}
\label{comparison}
\end{figure}

To compare the kinematics of the six halo CP2 stars to other nonpeculiar stars of similar spectral classes, we used the catalog of \citet{xiang_2022}, which lists stellar labels, radial velocities, and astrometric information for a large sample of hot stars. From this catalog, we only retained sources with relative parallax errors lower than 20\%, \texttt{RUWE}$<$1.4, and astrometric \texttt{fidelity}$>$0.5. We further excluded stars with unreliable radial velocities with nonpositive and high ($>$5~$\mathrm{km \, s^{-1}}$) radial velocity errors. Taking into account the typical physical characteristics of CP2 stars \citep[listed in, e.g.,][]{preston_1974}, we consider only stars within the temperature range 8000\,K\,$<$\,$T_{\mathrm{eff}}$\,$<$\,15000\,K and surface gravities of 3.9\,$<$\,$\log(g)$\,$<$\,4.4. This approximately corresponds to main-sequence stars of spectral classes B5 to A5, which corresponds to the main locus of CP2 stars. Following this method, a sample of $\sim$50\,000 stars (the nonpeculiar sample) was retained.

The vast majority of stars from the nonpeculiar sample are concentrated in the region near $\sim$(230, 0)~$\mathrm{km \, s^{-1}}$, which is consistent with the kinematics of the Galactic disk (see Fig.~\ref{comparison}). The sample also contains a small number of stars having halo-like kinematics that populate the same regions in the Toomre diagram as the halo CP2 stars identified in this study. However, these CP2 stars are well separated from the main thin and thick components of the Galactic disk, as delineated by the sample of nonpeculiar stars. In order to quantify this, we computed relative probabilities for the thick-disk-to-halo membership (TD/H) for the studied stars using the kinematic method described in \citet{bensby_2003}. All studied stars have TD/H$<$0.01, therefore their kinematics are more compatible with the halo than the thick disk. 

For each star identified as a halo candidate, we calculated its orbit in the Galaxy using the software package \textit{galpy}\footnote{\url{http://github.com/jobovy/galpy}} \citep{galpy}. We assumed the Galactic potential as given by the \textit{galpy} model \texttt{MWPotential2014}, which consists of a bulge, a Miyamoto-Nagai disk, and a dark matter halo component that is modeled by a Navarro-Frenk-White potential \citep{galpy}. The orbits were integrated back in time for 200 Myr, with a time step of 0.1~Myr. As in the previous section, we utilized the Monte Carlo approach, drawing 500 realizations of the orbit of each star based on the astrometric observables and their uncertainties.

All objects exhibit typical halo-like orbits and seem to have crossed the Galactic plane at least once within these 200 Myr, with most objects crossing it several times. The orbit of LAMOST J122746.05+113635.3 (Gaia DR3 3907547639444408064) is shown in Fig.~\ref{fig:orbit_star1} as an example. The orbits for the rest of the objects are shown in the Appendix in Figs.~\ref{fig:orbit_star_676074826449368704}--\ref{fig:orbit_star_1419257997205516416}.

The kinematic properties and the estimated times of the last crossing of the Galactic plane ($t_{\mathrm{cross}}$) for the studied objects are given in Table~\ref{table2_kinematics}. For all objects,  $t_{\mathrm{cross}}$ is much smaller than the typical lifetime of CP2 stars. This makes it difficult to  determine the likely exact birth site (i.e., cluster or association) and to identify the process that imparted runaway velocity to these objects. This is further exacerbated by the fact that the CP2 phase is commonly observed during the later stages of the main-sequence evolution. \citet{huemmerich_2020}
found a mean fractional age of 63\% among their sample. However,
 young CP2 stars also exist in open clusters \citep{poehnl05},
but they seem to be an exception. 
Therefore, we are dealing with relatively old objects, approaching the end of their main-sequence lifetime, which must have crossed the plane multiple times. It is therefore impossible to determine the exact ejection site or the ejection velocity with any degree of certainty for these objects. However, according to the computed orbits, the stars do not approach the Galactic center region (perhaps except for Gaia EDR3 676074826449368704, although with a very low probability) or seem to have an extragalactic origin. Most likely, they originate in the Galactic disk where they were imparted runaway velocities by some process, presumably through dynamical interactions \citep{1967BOTT....4...86P} or in the binary supernova scenario \citep{1961BAN....15..265B}. The former mechanism operates predominantly in dense and young star clusters and associations where runaway velocity can be imparted to a star through close stellar encounters. An example of this are close binary--binary encounters, which most likely result in the ejection of the least massive star involved in the interaction. In the binary supernova scenario, runaway velocity can be imparted to the secondary star of a binary system as a consequence of the rapid mass-loss and asymmetry in a core-collapse supernova explosion of the primary component, which often leads to the dissolution of the binary system. In this case the secondary leaves the system with approximately its previous orbital velocity.

\begin{figure*}[htp]

\centering
\includegraphics[width=.3\textwidth]{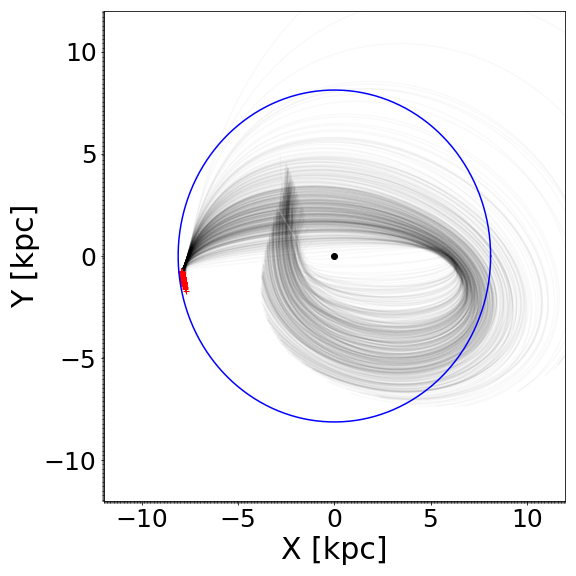}\hfill
\includegraphics[width=.3\textwidth]{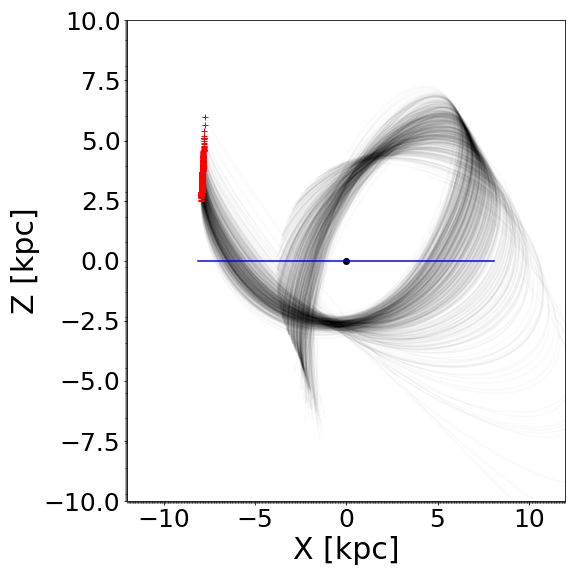}\hfill
\includegraphics[width=.3\textwidth]{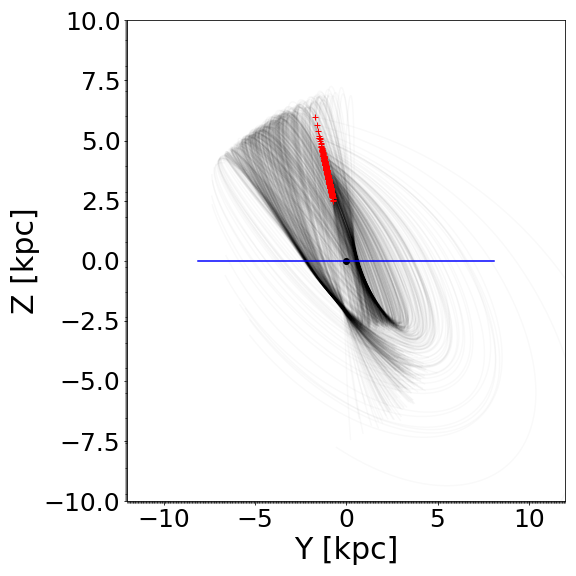}

\caption{Orbit of Gaia EDR3 3907547639444408064 in XYZ galactocentric coordinates, integrated 200 Myr backward in time. The red dots represent the current position of the star and the black dot represents the Galactic center. The thin gray lines show 500 orbits resulting from Monte Carlo realizations obtained from the astrometric parameters and their uncertainties (correlations between these parameters also taken into account) in order to illustrate the uncertainty in the derived orbits.}
\label{fig:orbit_star1}

\end{figure*}

\section{Summary and conclusions}
We investigated the possibility of the presence of a population of CP2 stars in the Galactic halo, as first proposed by \citet{huemmerich_2020}, who identified ten potential halo CP2 stars with significant height above or below the Galactic disk. Using the novel improved astrometry provided by \textit{Gaia} EDR3 \citep{gaia_edr3_summary,gaia_edr3_astrometry}, we studied these potential halo objects in more detail, supplementing this sample with a number of CP star candidates from the SDSS-based sample of Paunzen et al. (in prep.). 

The \textit{Gaia} astrometry of these objects was complemented by radial velocities from the LAMOST and SDSS surveys \citep{zhao_2012,cui_2012,blanton_2017}, yielding 6D information that can be used to ascertain the halo membership. From the initial sample of ten halo CP2 star candidates from \citet{huemmerich_2020}, only one star (LAMOST J122746.05+113635.3 = GAIA EDR3 3907547639444408064) shows kinematics consistent with a halo star classification.

However, we were able to identify an additional 14 CP star candidates from the SDSS sample with kinematics consistent with a halo membership. More detailed spectroscopic analysis revealed that seven of these stars are metal weak, which is typical for halo stars, and two further objects are chemically normal F dwarfs (although the noisiness of the corresponding spectrum precluded accurate spectral classification in the case of one of these stars). These nine objects, therefore, were excluded from further analysis. The remaining six objects were spectroscopically confirmed as CP2 stars.

In summary,  we identified six CP2 stars with kinematical properties in agreement with a halo membership. All these stars also display typical halo-like orbits. However, because of the relatively great age of the studied objects, it was impossible to accurately determine their specific point of origin or the type of process that has imparted runaway velocities to these objects. Orbits calculated backward for 200~Myr suggest that most of the objects crossed the Galactic plane several times in this timeframe and that there is no evidence for their origin in the vicinity of the Galactic center, suggesting that these CP2 stars probably belong to the population of runaway stars that were ejected through dynamical interactions or in the binary supernova scenario from the Galactic disk, making them the first runaway CP2 stars known. 

By analysis of ZTF and ATLAS data, we confirmed five of the CP2 stars as ACV variables. They exhibit photometric variability with periods and amplitudes typical for this class of objects. 

We showcase that the combination of precise astrometry from \textit{Gaia} and line-of-sight velocities from spectroscopic surveys such as LAMOST and SDSS can be used in a 6D survey to effectively study the Galactic kinematics of numerous peculiar types of objects. It can be expected that the sample of halo CP stars will be expanded in the coming years with future releases of the spectroscopic data and improved astrometry in \textit{Gaia} DR4.

\begin{acknowledgements}
This work has made use of data from the European Space Agency (ESA) mission
{\it Gaia} (\url{https://www.cosmos.esa.int/gaia}), processed by the {\it Gaia}
Data Processing and Analysis Consortium (DPAC,
\url{https://www.cosmos.esa.int/web/gaia/dpac/consortium}). Funding for the DPAC
has been provided by national institutions, in particular the institutions
participating in the {\it Gaia} Multilateral Agreement.
Guoshoujing Telescope (the Large Sky Area Multi-Object Fiber Spectroscopic Telescope LAMOST) is a National Major Scientific Project built by the Chinese Academy of Sciences. Funding for the project has been provided by the National Development and Reform Commission. LAMOST is operated and managed by the National Astronomical Observatories, Chinese Academy of Sciences. 
Funding for the Sloan Digital Sky 
Survey IV has been provided by the 
Alfred P. Sloan Foundation, the U.S. 
Department of Energy Office of 
Science, and the Participating 
Institutions. 

SDSS-IV acknowledges support and 
resources from the Center for High 
Performance Computing  at the 
University of Utah. The SDSS 
website is www.sdss.org.

SDSS-IV is managed by the 
Astrophysical Research Consortium 
for the Participating Institutions 
of the SDSS Collaboration including 
the Brazilian Participation Group, 
the Carnegie Institution for Science, 
Carnegie Mellon University, Center for 
Astrophysics | Harvard \& 
Smithsonian, the Chilean Participation 
Group, the French Participation Group, 
Instituto de Astrof\'isica de 
Canarias, The Johns Hopkins 
University, Kavli Institute for the 
Physics and Mathematics of the 
Universe (IPMU) / University of 
Tokyo, the Korean Participation Group, 
Lawrence Berkeley National Laboratory, 
Leibniz Institut f\"ur Astrophysik 
Potsdam (AIP),  Max-Planck-Institut 
f\"ur Astronomie (MPIA Heidelberg), 
Max-Planck-Institut f\"ur 
Astrophysik (MPA Garching), 
Max-Planck-Institut f\"ur 
Extraterrestrische Physik (MPE), 
National Astronomical Observatories of 
China, New Mexico State University, 
New York University, University of 
Notre Dame, Observat\'ario 
Nacional / MCTI, The Ohio State 
University, Pennsylvania State 
University, Shanghai 
Astronomical Observatory, United 
Kingdom Participation Group, 
Universidad Nacional Aut\'onoma 
de M\'exico, University of Arizona, 
University of Colorado Boulder, 
University of Oxford, University of 
Portsmouth, University of Utah, 
University of Virginia, University 
of Washington, University of 
Wisconsin, Vanderbilt University, 
and Yale University.

MP was supported by the European Regional Development Fund, 
project No. ITMS2014+: 313011W085.
\end{acknowledgements}

\bibliographystyle{aa} 
\bibliography{paper_bibliography.bib}
-----------------------------------------------------

\begin{appendix} 
\section{Phased light curves and Fourier spectra of the halo CP2 stars}

In this section, we present the phased light curves and Fourier spectra of the spectroscopically confirmed halo CP2 stars.

\begin{figure*} [h!]
\includegraphics[width=\columnwidth]{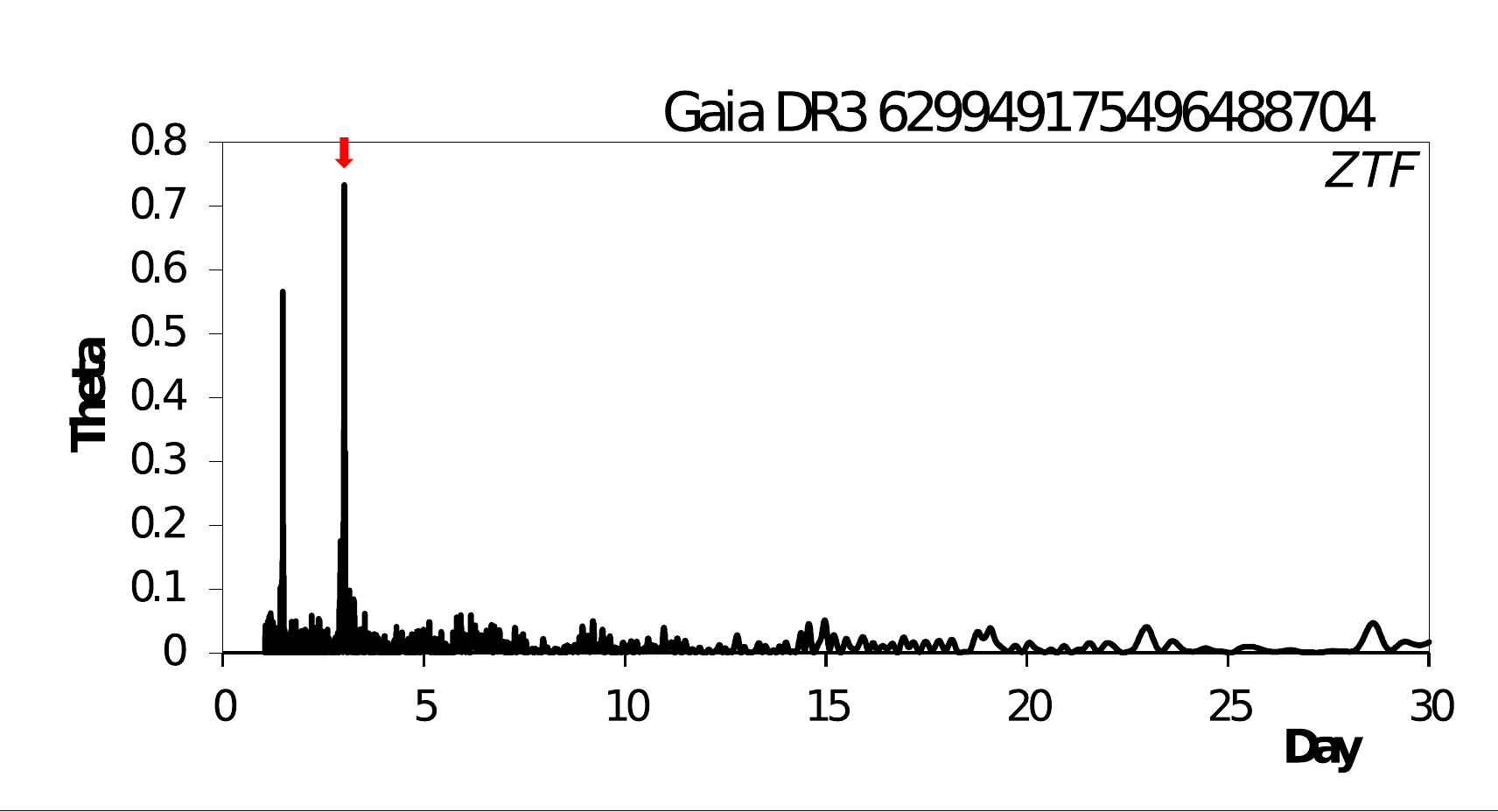}
\includegraphics[width=\columnwidth]{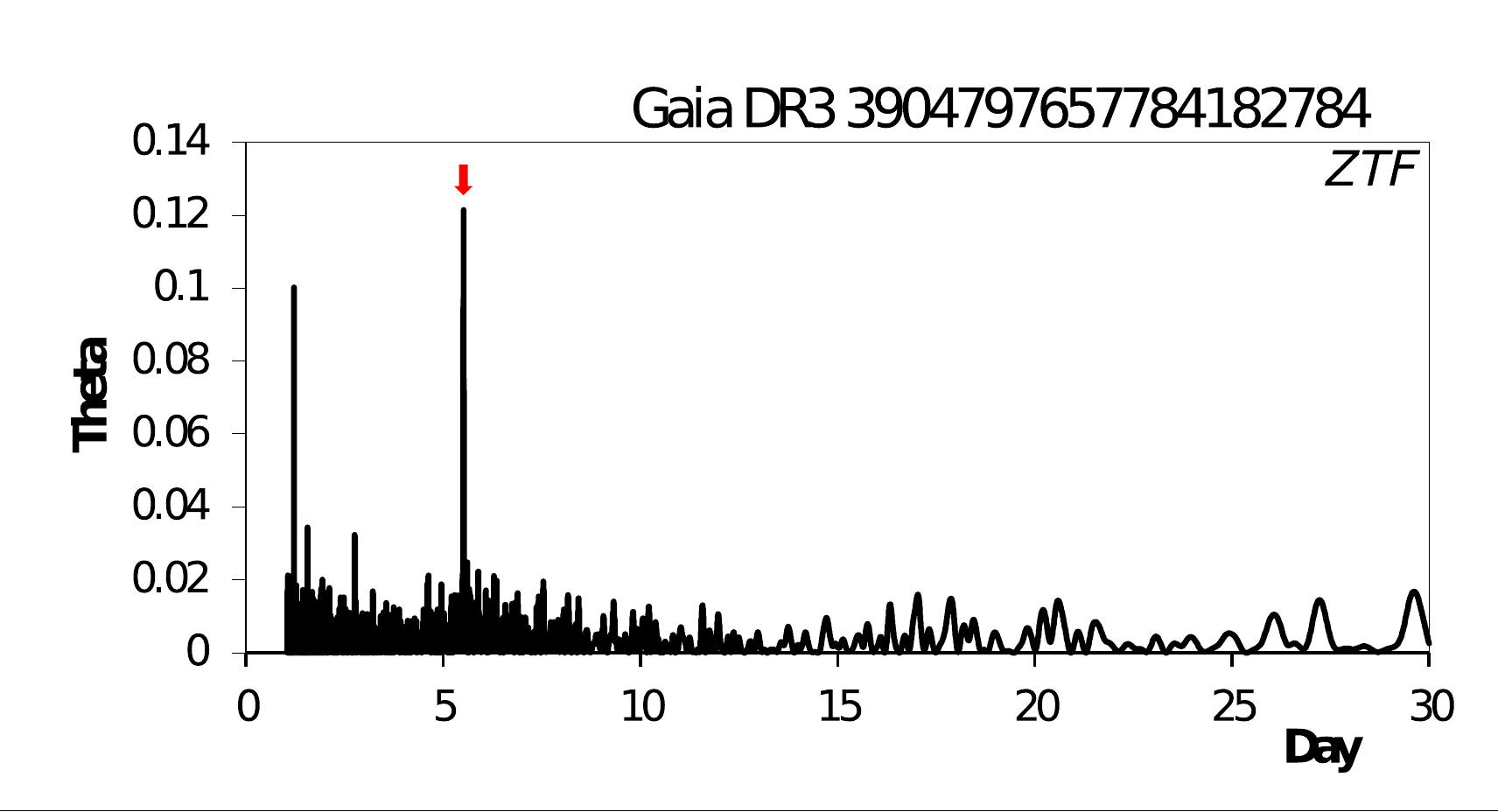}
\includegraphics[width=\columnwidth]{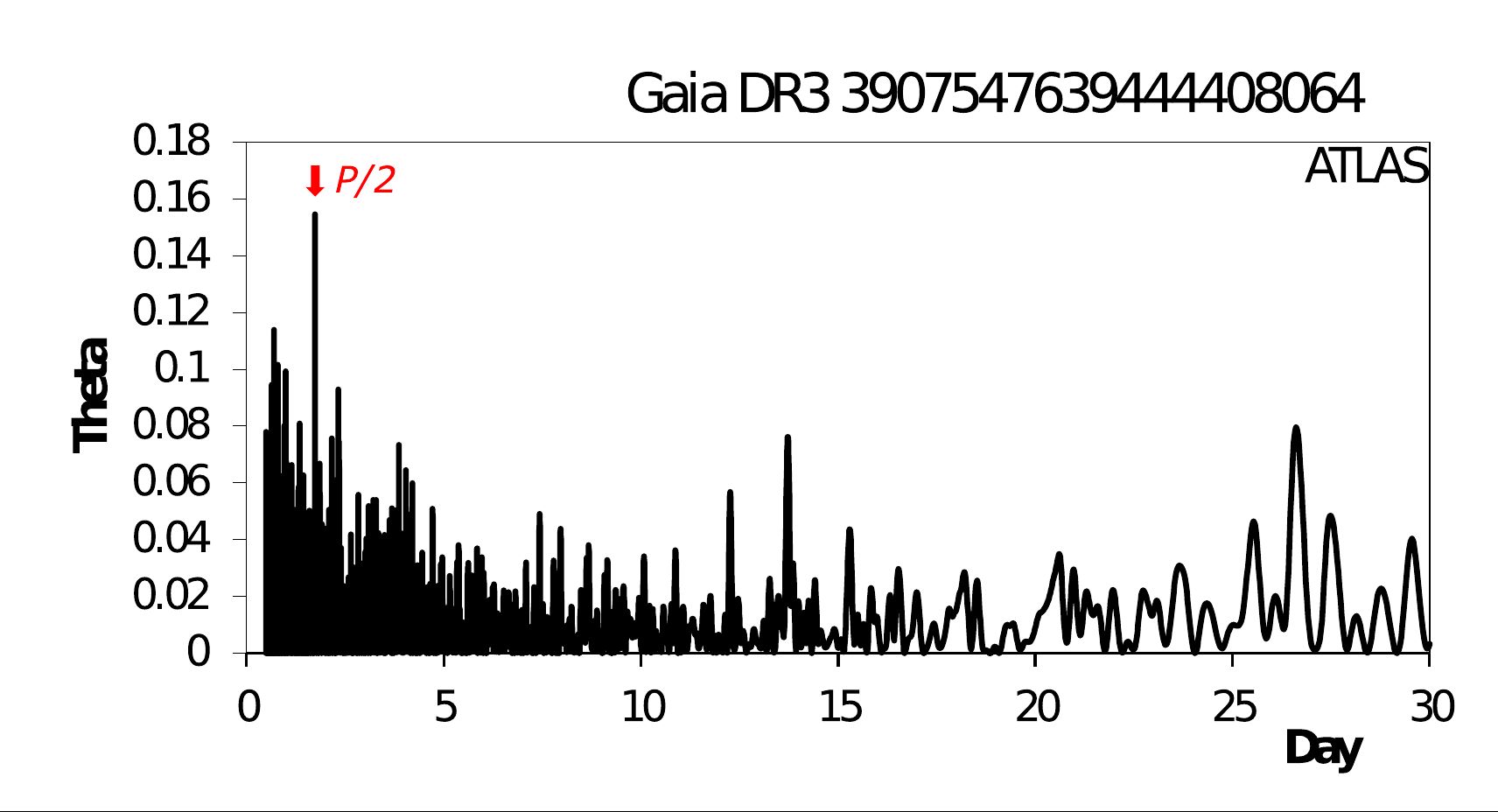}
\includegraphics[width=\columnwidth]{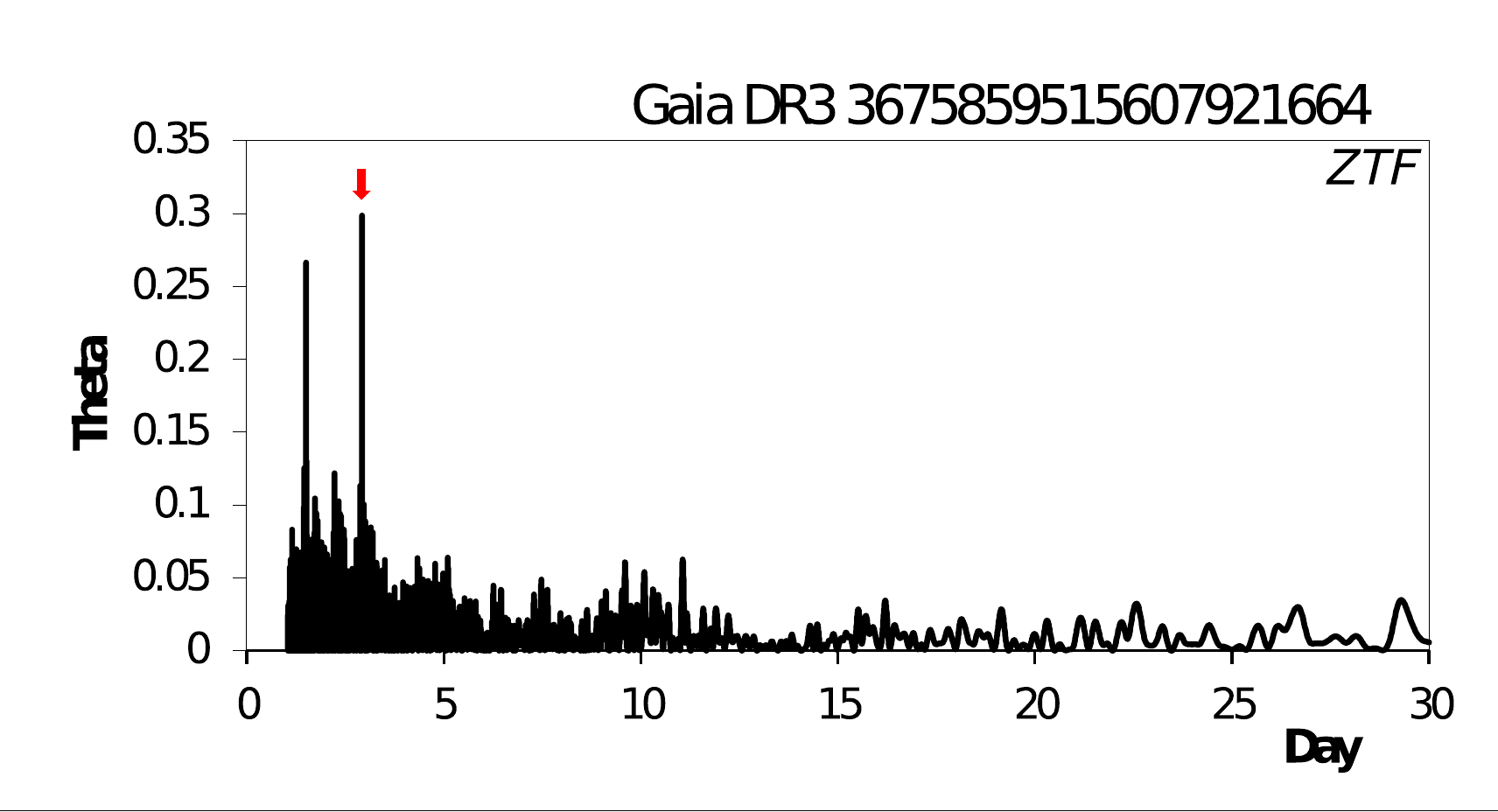}
\includegraphics[width=\columnwidth]{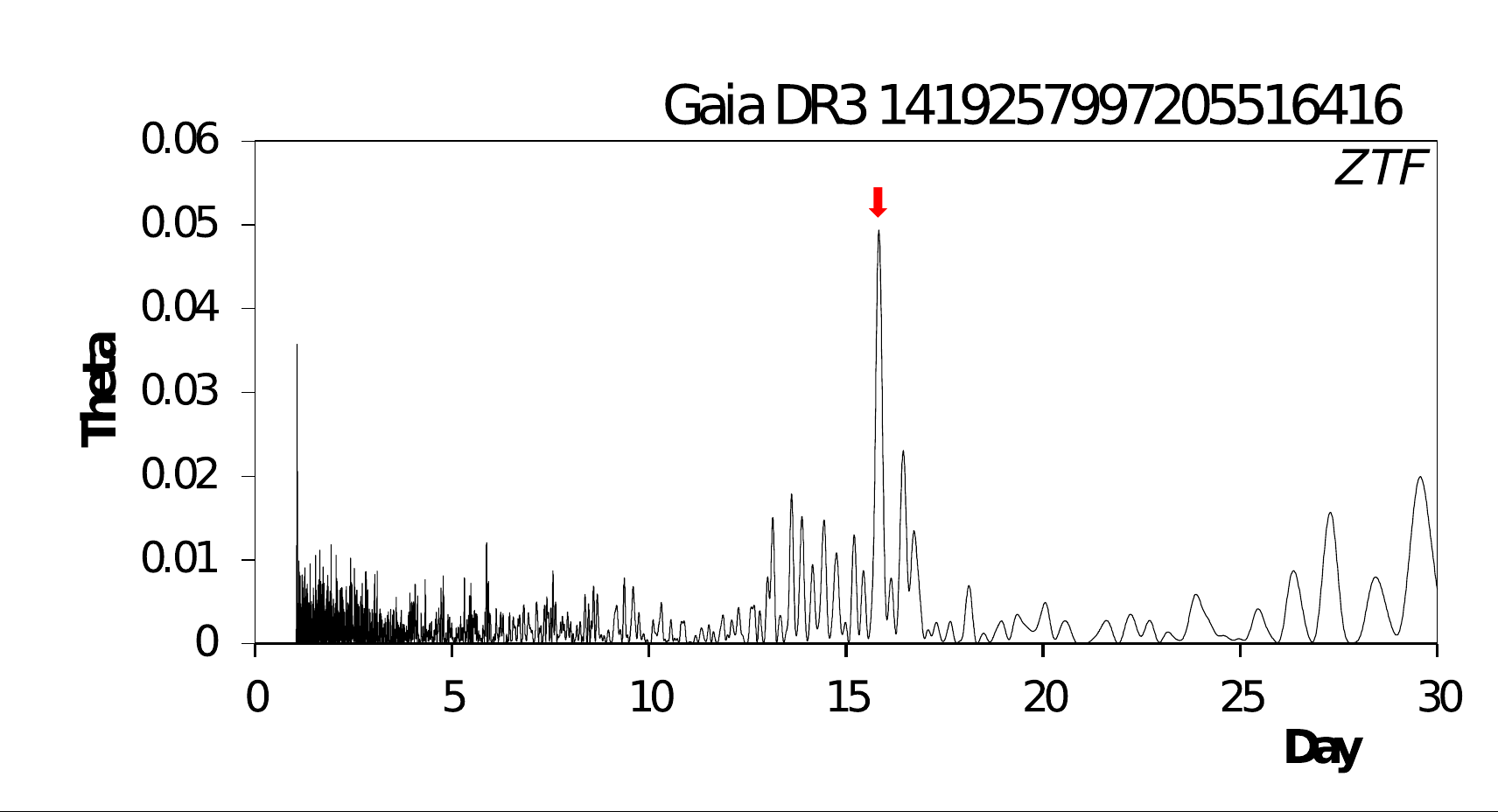}
\caption{Fourier spectra of the five photometrically variable CP2 stars. The data source (ZTF or ATLAS) is indicated in the upper right corner of each panel. In the case of Gaia DR3 3907547639444408064, the highest peak corresponds to half the true rotational period. For all other objects, the rotational periods correspond to the highest peaks (indicated by the red arrow in each panel).}
\label{fig_Fourier_spectra}
\end{figure*}

\begin{figure*}
\includegraphics[width=\columnwidth]{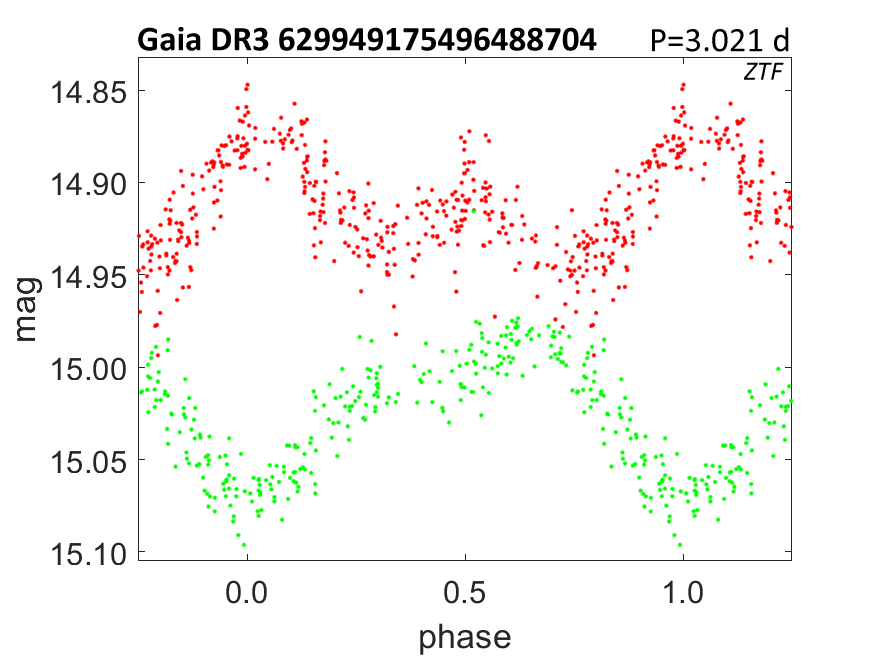}
\includegraphics[width=\columnwidth]{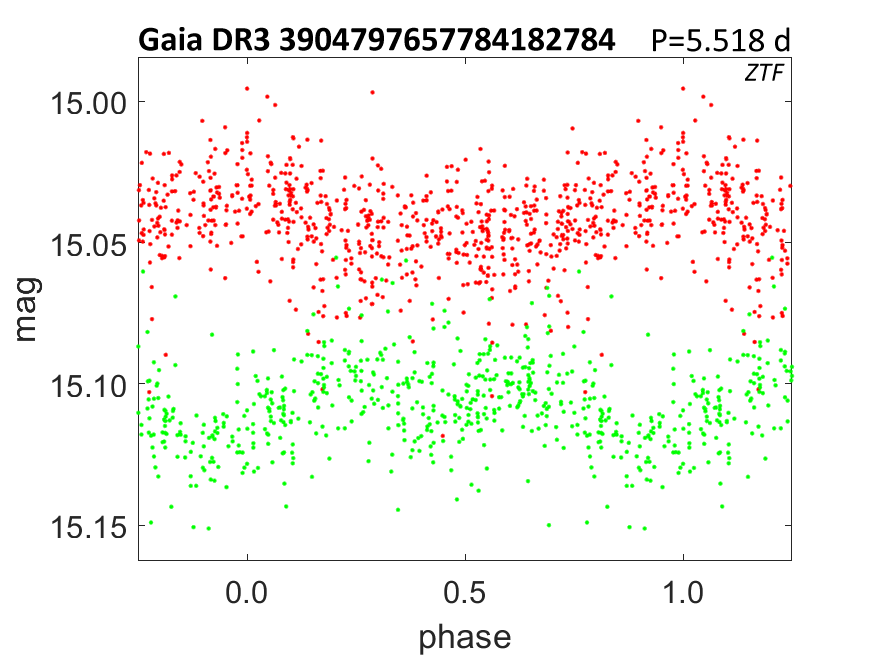}
\includegraphics[width=\columnwidth]{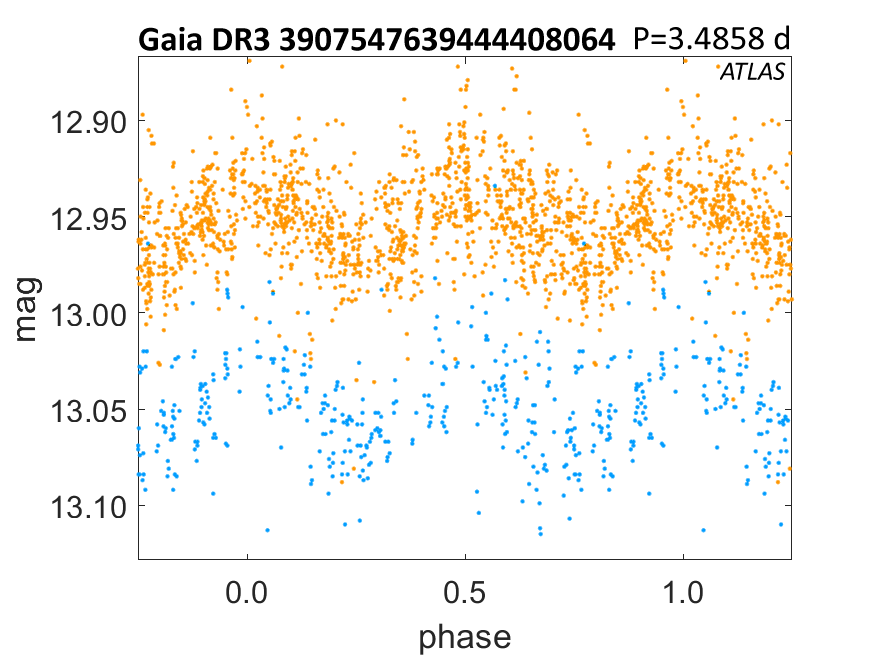}
\includegraphics[width=\columnwidth]{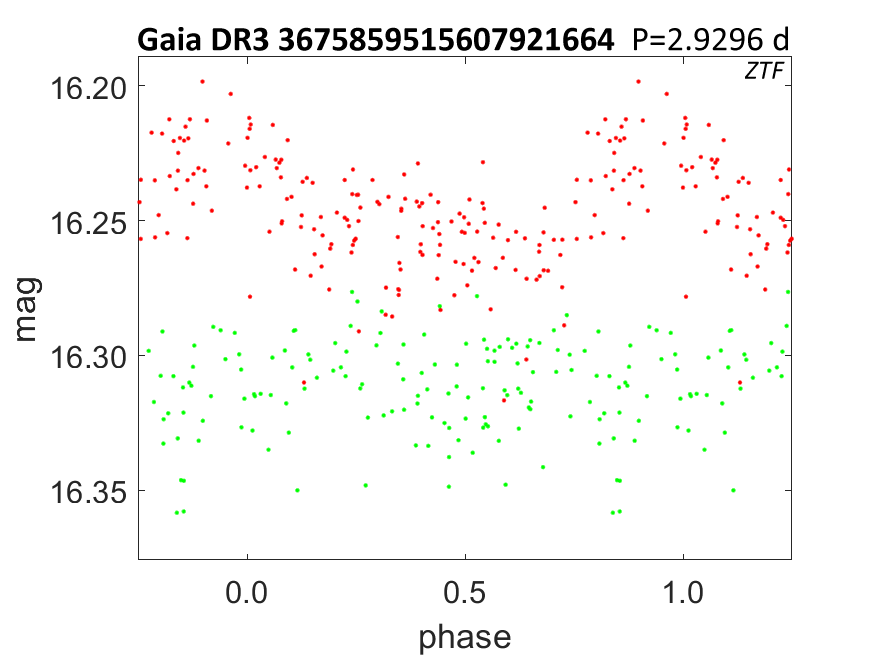}
\includegraphics[width=\columnwidth]{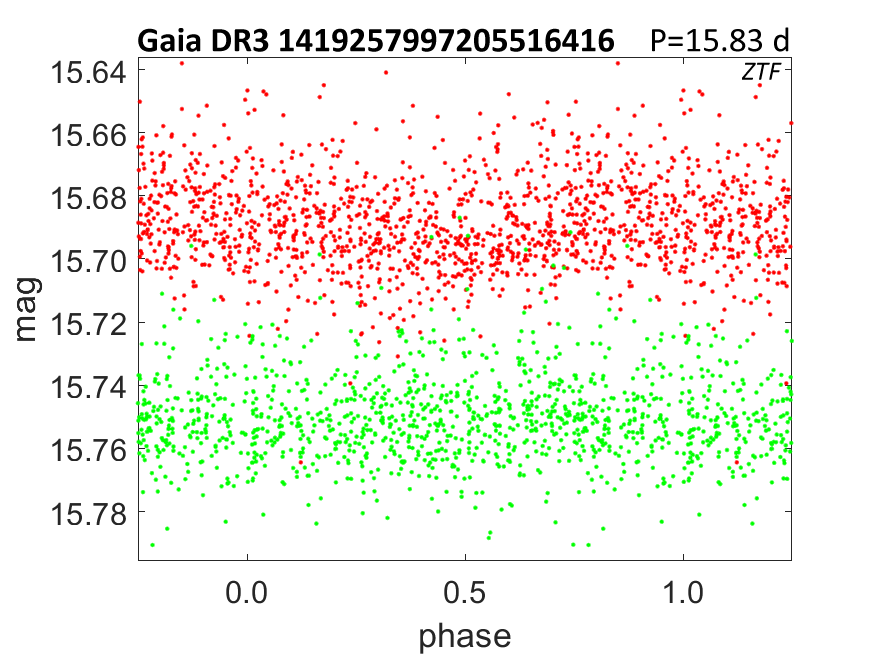} 
\caption{Phased light curves of the five photometrically variable CP2 stars. The data source (ZTF or ATLAS) is indicated in the upper right corner of each panel. The red and green dots denote ZTF $r$ and $g$ band data, respectively. The orange and blue dots denote ATLAS $o$ and $c$ band data, respectively. The adopted rotational periods are also indicated.}
\label{fig_phase_plots}
\end{figure*}

\section{Galactic orbits of the halo CP stars}
In this section we present the Galactic orbits for the rest of the CP stars from Table~\ref{table2_kinematics}.

\begin{figure*}[htp]
\centering
\includegraphics[width=.3\textwidth]{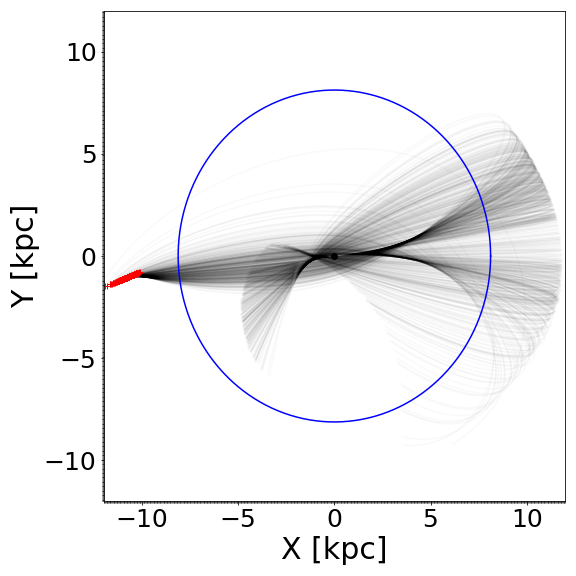}\hfill
\includegraphics[width=.3\textwidth]{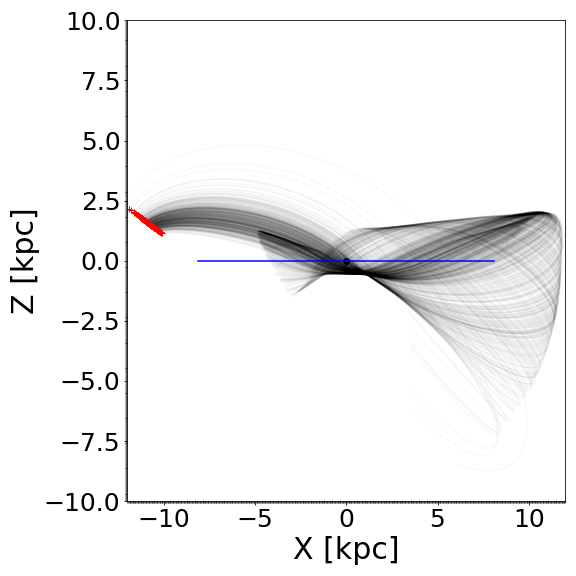}\hfill
\includegraphics[width=.3\textwidth]{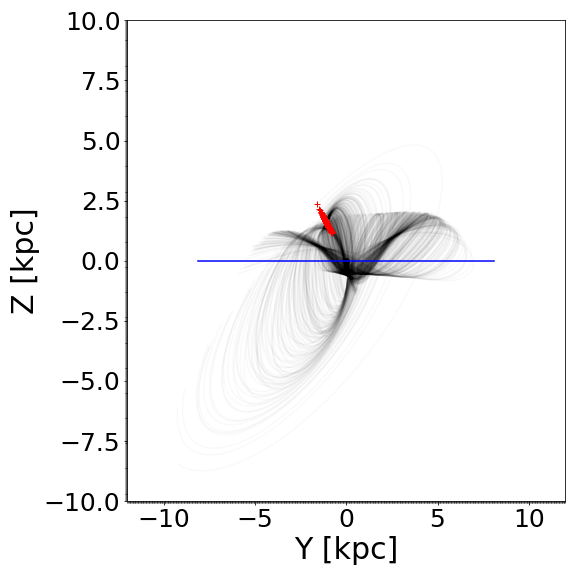}

\caption{Same as Fig.~\ref{fig:orbit_star1}, but for Gaia DR3 676074826449368704.}
\label{fig:orbit_star_676074826449368704}
\end{figure*}

\begin{figure*}[htp]
\centering
\includegraphics[width=.3\textwidth]{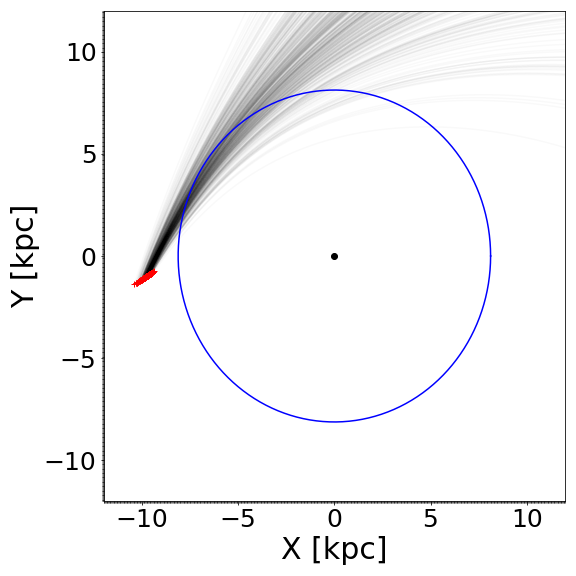}\hfill
\includegraphics[width=.3\textwidth]{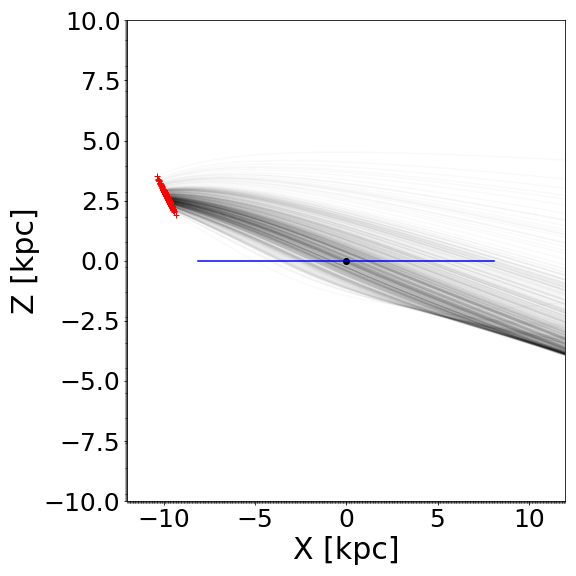}\hfill
\includegraphics[width=.3\textwidth]{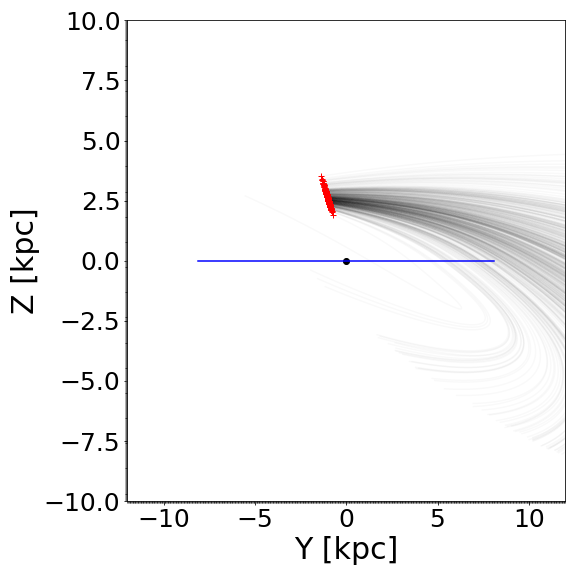}

\caption{Same as Fig.~\ref{fig:orbit_star1}, but for Gaia DR3 629949175496488704.}
\label{fig:orbit_star_629949175496488704}
\end{figure*}

\begin{figure*}[htp]
\centering
\includegraphics[width=.3\textwidth]{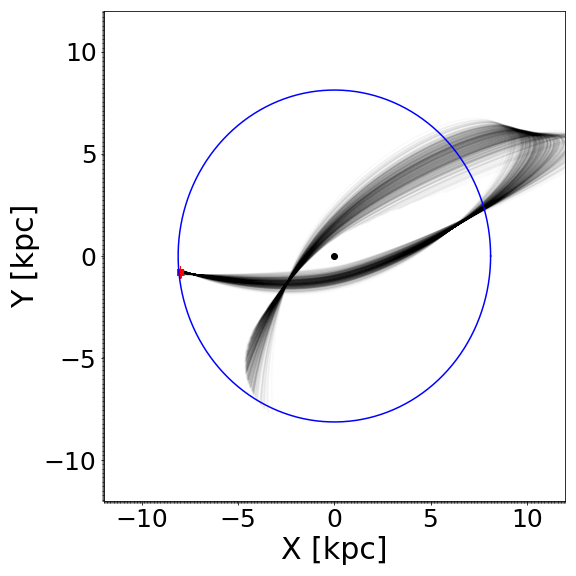}\hfill
\includegraphics[width=.3\textwidth]{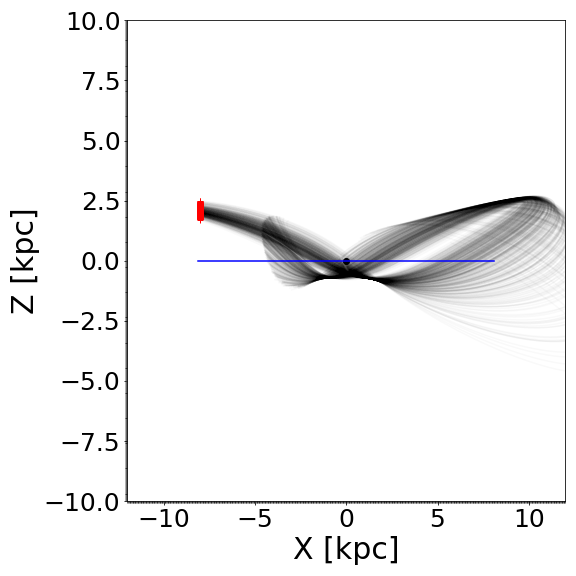}\hfill
\includegraphics[width=.3\textwidth]{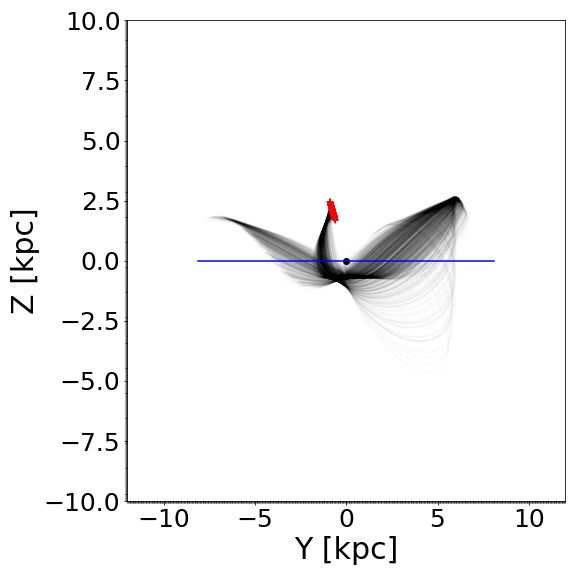}

\caption{Same as Fig.~\ref{fig:orbit_star1}, but for Gaia DR3 3904797657784182784.}
\label{fig:orbit_star_3904797657784182784}
\end{figure*}

\begin{figure*}[htp]
\centering
\includegraphics[width=.3\textwidth]{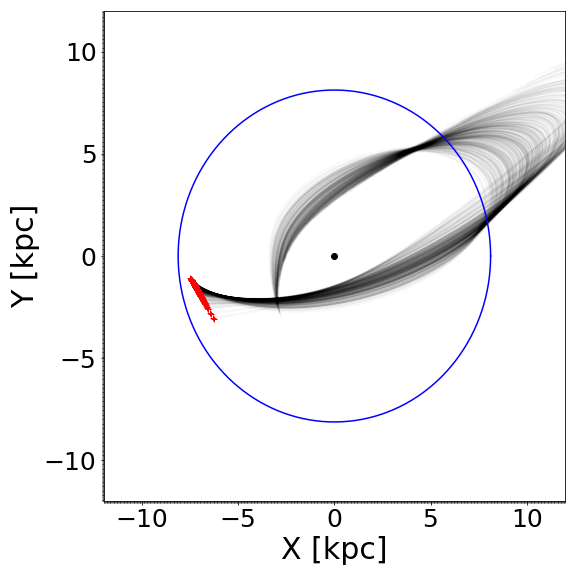}\hfill
\includegraphics[width=.3\textwidth]{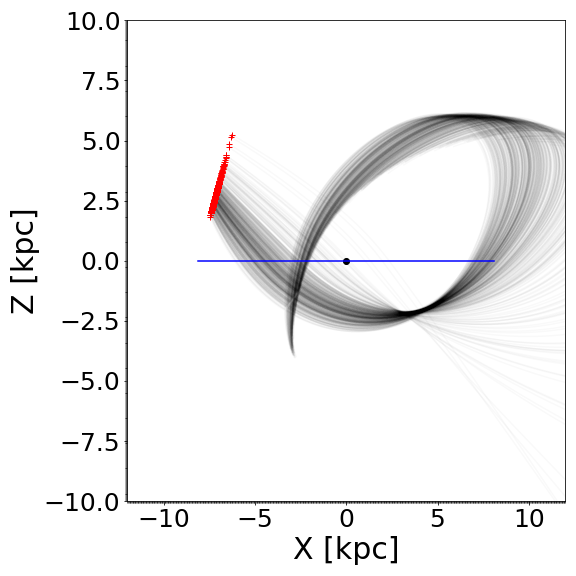}\hfill
\includegraphics[width=.3\textwidth]{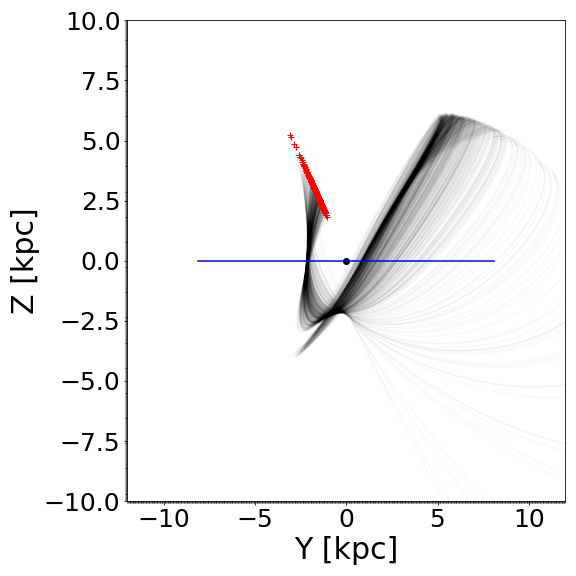}

\caption{Same as Fig.~\ref{fig:orbit_star1}, but for Gaia DR3 3675859515607921664.}
\label{fig:orbit_star_3675859515607921664}
\end{figure*}

\begin{figure*}[htp]
\centering
\includegraphics[width=.3\textwidth]{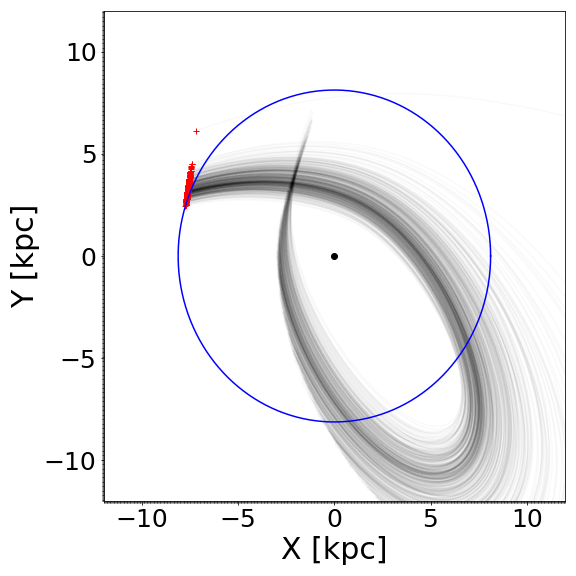}\hfill
\includegraphics[width=.3\textwidth]{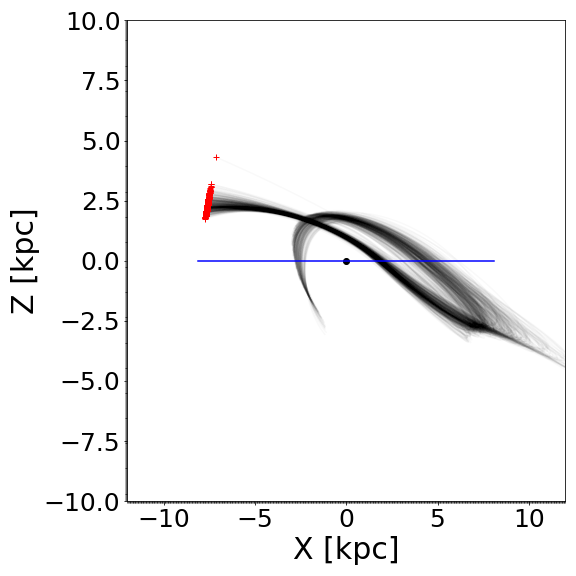}\hfill
\includegraphics[width=.3\textwidth]{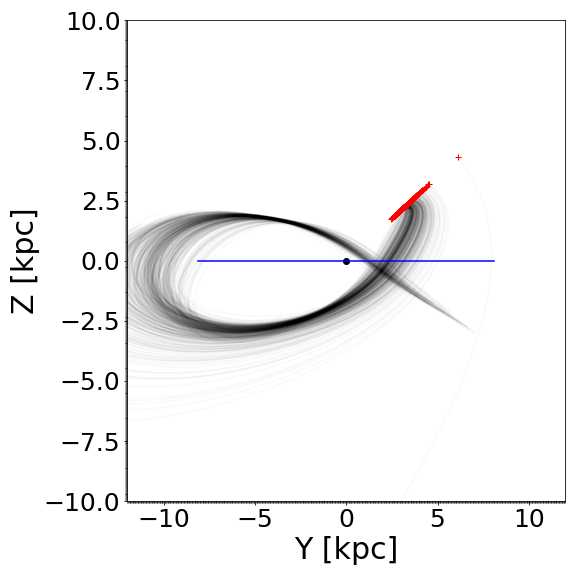}

\caption{Same as Fig.~\ref{fig:orbit_star1}, but for Gaia DR3 1419257997205516416.}
\label{fig:orbit_star_1419257997205516416}
\end{figure*}

\end{appendix}

\end{document}